\documentclass[conference]{IEEEtran}
\IEEEoverridecommandlockouts
\usepackage{cite}
\usepackage{amsmath,amssymb,amsfonts}
\usepackage{algorithmic}
\usepackage{graphicx}
\usepackage{textcomp}
\usepackage{xcolor}
\usepackage[utf8]{inputenc}
\usepackage{textgreek}
\usepackage{booktabs}

\DeclareUnicodeCharacter{03B2}{\textbeta}
\def\BibTeX{{\rm B\kern-.05em{\sc i\kern-.025em b}\kern-.08em
    T\kern-.1667em\lower.7ex\hbox{E}\kern-.125emX}}
\begin{document}

\title{Domain-Aware Geometric Multimodal Learning for Multi-Domain Protein-Ligand Affinity Prediction\\
\thanks{This work was partially supported by BlueBEAR and Baskerville, funded by the EPSRC and UKRI through the World Class Labs scheme (EP/T022221/1) and the Digital Research Infrastructure programme (EP/W032244/1) and operated by Advanced Research Computing at the University of Birmingham.}
}

\author{\IEEEauthorblockN{Shuo Zhang}
\IEEEauthorblockA{\textit{School of Computer Science} \\
\textit{University of Birmingham}\\
Birmingham, UK \\
sxz325@student.bham.ac.uk}
~\\
\and
\IEEEauthorblockN{Jian K. Liu*}
\IEEEauthorblockA{\textit{School of Computer Science} \\
\textit{University of Birmingham}\\
Birmingham, UK \\
j.liu.22@bham.ac.uk}
*Corresponding author
~\\
}

\maketitle

\begin{abstract}
The accurate prediction of protein-ligand binding affinity is important for drug discovery yet remains challenging for multi-domain proteins, where inter-domain dynamics and flexible linkers govern molecular recognition. Current geometric deep learning methods typically treat proteins as monolithic graphs, failing to capture the distinct geometric and energetic signals at domain interfaces. To address this, we introduce DAGML (Domain-Aware Geometric Multimodal Learning), a hierarchical framework that explicitly models domain modularity. DAGML integrates a pre-trained protein language model with a novel domain-aware geometric encoder to distinguish intra- and inter-domain features, while a motif-centric ligand encoder captures pharmacophoric compatibility. We further curate a specialized multi-domain affinity benchmark, classifying complexes by binding topology (e.g., interface vs linker binders). Extensive experiments demonstrate that DAGML achieves a 21\% reduction in MSE and a Pearson correlation of 0.726 compared to strong baselines. Ablation studies reveal that explicit modeling of domain interfaces is the primary driver of this improvement, particularly for ligands binding in the clefts between structural units. The code is available at https://github.com/jiankliu/DAGML.
\end{abstract}

\begin{IEEEkeywords}
Protein-Ligand Affinity Prediction, Multi-Domain Proteins, Multimodal Learning, Drug Discovery
\end{IEEEkeywords}

\begin{figure*}[htbp]
\centering
\includegraphics[width=0.95\textwidth]{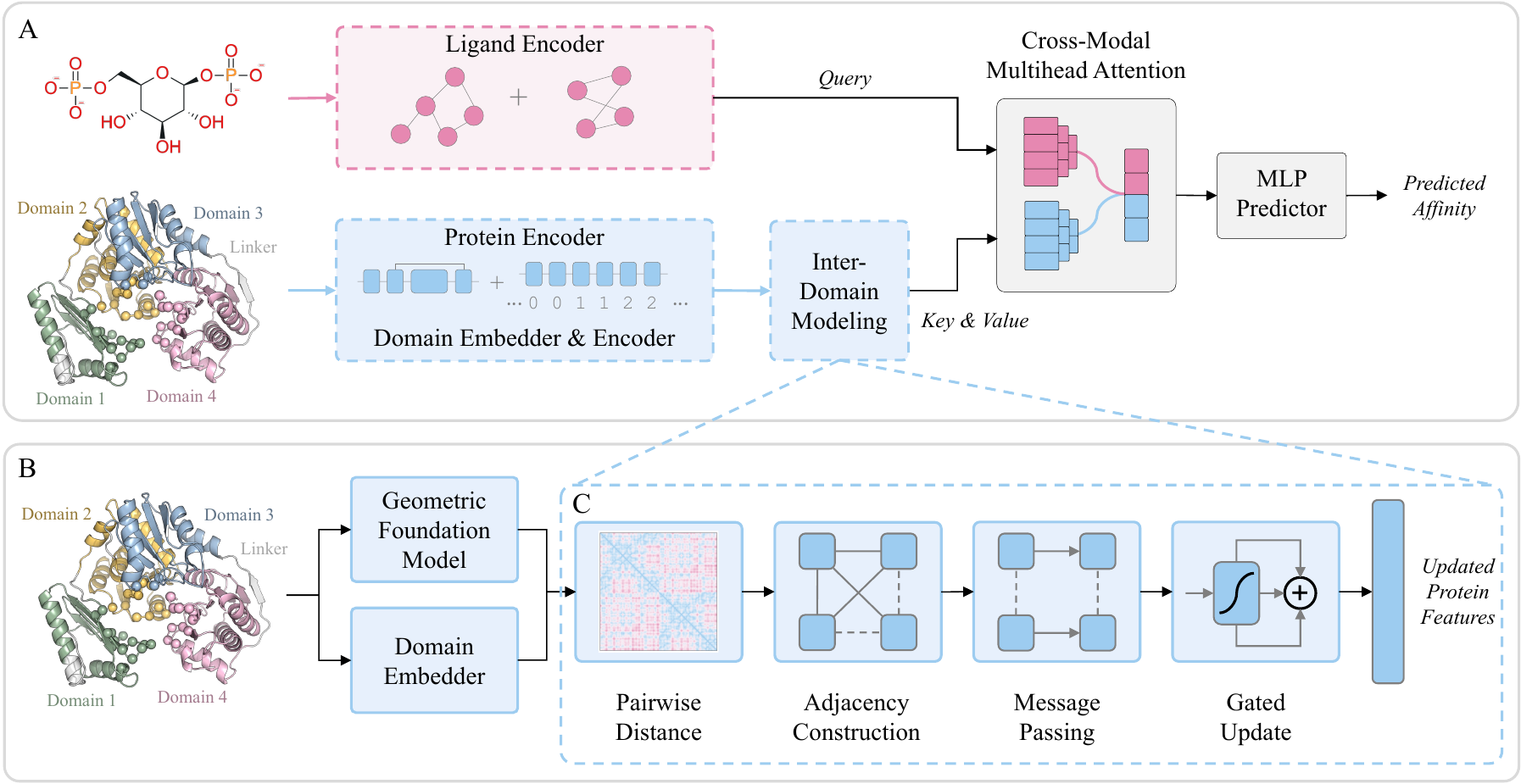}
\caption{The DAGML framework. (A) Overview of the complete affinity prediction pipeline. The model processes ligand and protein inputs through respective encoders, followed by an Inter-Domain Modeling module and a Cross-Modal Multihead Attention mechanism that feeds into an MLP Predictor. (B) The protein feature processing stage. (C) Detailed Inter-Domain Modeling module.}
\label{fig:model_arch}
\end{figure*}

\section{Introduction}
Precise prediction of protein-ligand binding affinity is a cornerstone of rational drug discovery, essential for prioritizing lead compounds and reducing clinical trial attrition~\cite{stark2022equibind}. While recent advances in geometric deep learning and protein language models (PLMs) have revolutionized molecular representation~\cite{zhang2022esmgearnet, zhang2022gearnet, chenHybridGCNProteinSolubility2023, Lin2023esm2, zhangSeqProFTSequenceonlyProtein2025, zhouProCeSaContrastEnhancedStructureAware2025}, the majority of current approaches simplify protein targets into monolithic geometric entities, overlooking a fundamental biological reality: a significant proportion of the eukaryotic proteome consists of multi-domain proteins, where distinct structural units are tethered by flexible linkers~\cite{Gianni2025folding}. These complex architectures govern essential cellular processes through intricate inter-domain communications and large-scale conformational changes, often serving as important drug targets for complex pathologies~\cite{Basciu2025toward}.

Despite their prevalence, computational methods struggle to capture the energetic and structural nuances of these systems. Standard Graph Neural Networks (GNNs), while effective for single-domain representations, suffer from "over-smoothing" when applied to large, multi-domain graphs~\cite{chen2020measuring, su2024saprot}. In such architectures, message passing across the entire protein graph tends to dilute the distinct geometric signals. This dilution is particularly problematic for signals emerging from domain interfaces. These regions are frequently implicated in allosteric regulation and ligand binding~\cite{Sidhanta2023comparative, Weinstein2011correlation}. Furthermore, existing docking and affinity prediction models often fail to account for the plasticity of domain-domain interactions, which can range from permanent, rigid interfaces to transient, flexible associations~\cite{Sidhanta2023comparative}. Neglecting these specific structural dynamics limits the identification of cryptic or allosteric binding pockets within inter-domain clefts~\cite{Basciu2025toward}.

To address these limitations, we introduce DAGML (Domain-Aware Geometric Multimodal Learning), a unified framework designed to explicitly model the hierarchical organization of multi-domain proteins for affinity prediction. Motivated by the observation that domain tethering alters the energetic landscape and flexibility of constituent domains~\cite{Vishwanath2018}, DAGML moves beyond the monolithic treatment of protein structures. Our approach integrates a pre-trained protein language model to capture evolutionary semantics with a novel domain-aware geometric encoder that explicitly models inter-domain interfaces. By fusing this hierarchical protein representation with a motif-centric ligand encoding via cross-modal attention, DAGML captures the physicochemical compatibility between specific pharmacophores and dynamic protein clefts. This work bridges the gap between static structural representation and the dynamic modularity of biological systems, offering a robust tool for targeting complex multi-domain proteins.

\section{Methodology}
\label{sec:methodology}
We introduce DAGML, a geometric deep learning framework designed to predict the binding affinity of ligands to multi-domain proteins. Our approach addresses the limitations of standard graph neural networks (GNNs), which treat proteins as monolithic geometric entities, thereby failing to capture the specific structural dynamics of domain interfaces. As illustrated in Figure~\ref{fig:model_arch}, DAGML integrates a pre-trained protein language model with a domain-aware geometric encoder, fused with a motif-centric ligand representation via cross-modal attention.

\subsection{Hierarchical Protein Representation}
To accurately model multi-domain proteins, we require a representation that captures both evolutionary semantics and local geometric fine-structure. We represent the protein as a spatial geometric graph $\mathcal{G}_P=(\mathcal{V,E})$, where nodes $v_i\in\mathcal{V}$ represent amino acid residues located at their $\alpha$-carbon coordinates $\mathbf{x}_i\in\mathbb{R}^3$.

\textbf{Sequence-Structure Initialization}. We initialize node features using a pre-trained Protein Language Model (PLM) to capture evolutionary constraints that are not immediately obvious from geometry alone. Specifically, we use ESM-GearNet~\cite{zhang2022esmgearnet} which consists of ESM-2\cite{Lin2023esm2} and GearNet~\cite{zhang2022gearnet}. ESM-2 extracts residue-level embeddings $\mathbf{h}_i^{(0)}\in\mathbb{R}^d$, where $d$ denotes the hidden dimension of the language model's final layer. GearNet encodes local geometry by constructing a multi-relational graph using sequential, spatial (radius), and K-nearest neighbor (KNN) edges. The initial structural encoding $\mathbf{h}_i^{(struct)}$ is obtained via relational message passing over these local edges.

\textbf{Domain-Awareness Injection}. Standard geometric encoders are agnostic to the high-level topological organization of proteins. To remedy this, we introduce explicit domain semantics. Let $\mathcal{D}=\{d_1,d_2,...,d_K\}$ denote the set of structural domains identified within the protein chain. We assign a domain indicator $\delta (i)\in \{0,...,K\}$ to each residue $v_i$, where $\delta(i)=0$ denotes a flexible linker or unstructured region, and $\delta(i)>0$ denotes a specific structured domain. We learn a domain embedding $\mathbf{e}_{dom}\in \mathbb{R}^d$ for each unique domain index and inject it into the node features via a residual connection:
\begin{equation}
    \mathbf{h}_i=\mathbf{h}_i^{(struct)}+\mathbf{e}_{dom}(\delta(i))
\end{equation}

This injection breaks the permutation invariance regarding domain identity, allowing the network to distinguish between catalytic and regulatory domains even if they share similar local geometries.

\begin{figure*}[]
\centering
\includegraphics[width=1.0\textwidth]{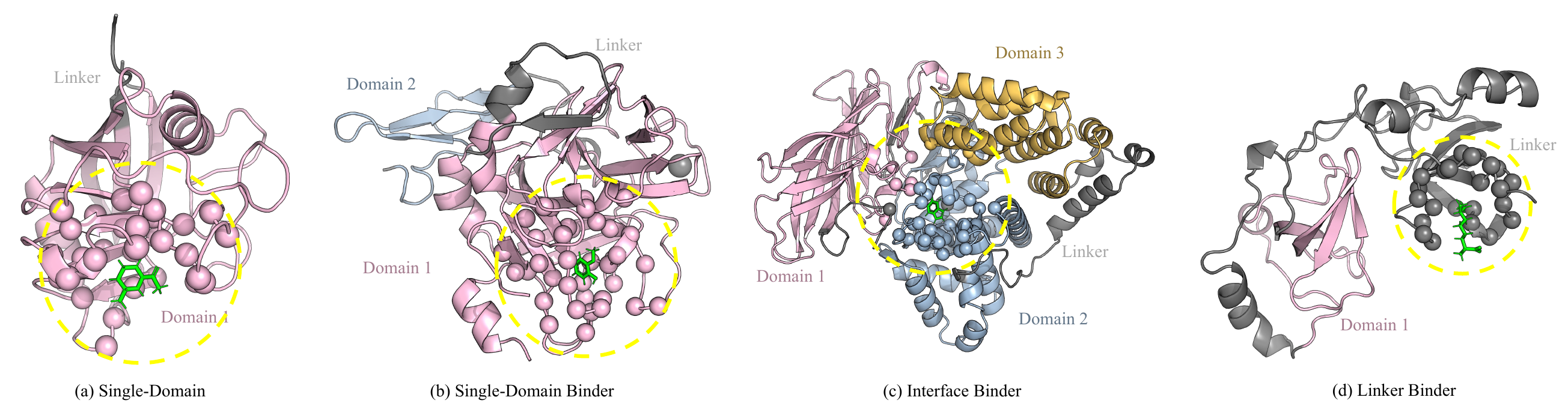}
\caption{Examples of geometric interface classifications. (a) Single-Domain: A ligand binds to a protein consisting of one domain. (b) Single-Domain Binder: In a multi-domain architecture, the ligand interacts only with residues of a single domain. (c) Interface Binder: The ligand occupies the cleft between two distinct domains. (d) Linker Binder: The ligand interacts only with residues within a flexible linker region.}
\label{fig:domain_example_flat}
\end{figure*}

\subsection{Modeling Inter-Domain Interfaces}
\label{sec:modelinginterdomaininterfaces}
A core inductive bias of our method is that allosteric regulation and ligand binding in multi-domain proteins frequently occur at domain-interfaces where clefts are formed between two distinct structural units. Standard GNNs tend to over-smooth features across the entire graph, diluting the signal from these critical regions. We propose a sparse Inter-Domain Message Passing mechanism to explicitly model these interfaces while filtering out geometric noise from flexible linkers.

We construct a sparse set of inter-domain edges, $\mathcal{E}_{inter}$, where an edge $e_{ij}$ exists between residues $v_i$ and $v_j$ if and only if three conditions are met:
\begin{enumerate}
    \item \textbf{}{Spatial Proximity}: $\parallel \mathbf{x}_i-\mathbf{x}_j\parallel<\tau$, where $\tau$ is a distance cutoff (e.g., 8\r{A}).
    \item \textbf{}{Domain Distinctness}: The residues belong to different domains: $\delta(i)\neq\delta(j)$.
    \item \textbf{}{Structural Validity}: Neither residue belongs to a flexible linker: $\delta(i)\neq0\bigwedge \delta(j)\neq0$.
\end{enumerate}

This construction effectively creates a structural interface graph that isolates the rigid clefts formed between domains. We then perform a message passing update specifically on this edge set:
\begin{gather}
    \mathbf{m}_{i\leftarrow j}=\phi_{msg}(\mathbf{h}_j), \forall j\in \mathcal{N}_{inter}(i)\\
    \mathbf{h}'_i =\mathbf{h}_i+\phi_{update}(\mathbf{h}_i,\sum_{j\in\mathcal{N}_{inter}(i)}\frac{1}{\sqrt{d_id_j}}\mathbf{m}_{i\leftarrow j})
\end{gather}
where $\mathcal{N}_{inter}(i)$ denotes the neighbors of $i$ in $\mathcal{E}_{inter}$, and $\phi_{msg},\phi_{update}$ are learnable neural networks. This layer enriches the representations of interface residues with features from their spatial neighbors in opposing domains, effectively highlighting potential allosteric pockets for the subsequence readout phase.

\subsection{Motif-Centric Ligand Encoding}
Traditional ligand encoders often treat molecules as flat graphs of atoms, potentially obscuring the pharmacophoric features (motifs) that drive specific binding. To address this, we utilize Molecular Motif Learning (MotiL)~\cite{Liu2025motil}. MotiL decomposes the ligand $\mathcal{G}_L$ into a set of chemically meaningful subgraphs (motifs) and performs bi-scaled pre-training to align global molecular semantics with local functional group properties.

We extract the global ligand embedding $\mathbf{H}_L\in\mathbb{R}^{d_{ligand}}$ from the pre-trained MotiL encoder. This vector serves as a learned latent fingerprint that is motif-aware, mathematically encoding the presence of functional groups (e.g., rings, polar groups) and the global scaffold topology. This ensures the representation remains highly sensitive to the specific chemical properties that determine binding affinity.

\subsection{Cross-Modal Attention and Binding Prediction}
To predict the binding affinity, we must identify which specific protein residues interact with the ligand motifs. We employ a Cross-Modal Attention mechanism where the ligand motifs query the domain-aware protein residues.

Let $\mathbf{Q}=\mathbf{H}_L\mathbf{W}_Q$ be the query matrix derived from the ligand features, and $\mathbf{K}=\mathbf{H}_P\mathbf{W}_K, \mathbf{V}=\mathbf{H}_P\mathbf{W}_V$ be the key and value matrices derived from the updated protein residues $\mathbf{H}_P={\mathbf{h}'_1,...,\mathbf{h}_N'}$. We compute the attention matrix $\alpha$:
\begin{equation}
    \alpha=\text{softmax}(\frac{\mathbf{Q}\mathbf{K}^\intercal)}{\sqrt{d_k}}
\end{equation}

The interaction context vector $\mathbf{c}$ is obtained by aggregating the protein features weighted by their relevance to the ligand features:
\begin{equation}
    \mathbf{c}=\alpha \mathbf{V}
\end{equation}

Finally, the context vector $\mathbf{c}$ is passed through a Multi-Layer Perceptron (MLP) to predict the scalar binding affinity. This architecture ensures the model focuses on the specific protein regions enriched by our inter-domain message passing. By doing so, it prioritizes residues that are most chemically compatible with the ligand's functional motifs.

\section{Data Curation and Construction}
To evaluate the capacity of DAGML in capturing inter-domain allostery and interface binding, we constructed a specialized dataset derived from the PDBbind database. Unlike standard benchmarks that treat proteins as monolithic entities, our pipeline explicitly annotates domain boundaries and classifies ligand binding modes based on geometric proximity to specific structural units.

\subsection{Source Data and Splitting Strategy}
We utilized the PDBbind database~\cite{Liu2017pdbbind}, a comprehensive collection of protein-ligand complexes with experimentally measured binding affinities. To ensure a realistic evaluation of generalization capability, we adopted the time-based splitting strategy used in ~\cite{stark2022equibind}. The dataset was partitioned based on the release year of the complex structures: complexes released in 2019 or later served as the test set, while those released prior to 2019 constituted the training and validation sets. This temporal split prevents data leakage that often occurs in random splits due to the high sequence similarity of homologous proteins deposited over time~\cite{Joeres2025datasplitting}.

\subsection{Domain Annotation and Geometric Classification}
\label{sec:domain_annotation}
A core contribution of this work is the enrichment of PDBbind structural data with explicit domain semantics. We developed an automated pipeline to map and classify interaction interfaces within multi-domain architectures.

\textbf{Domain Mapping}: For every protein chain in the dataset, we mapped residue coordinates to structural families using the Pfam database~\cite{Mistry2020pfam}. Residues falling within defined Pfam boundaries were assigned specific domain identifiers, while residues falling outside these boundaries were annotated as flexible linkers or disordered regions.

\textbf{Geometric Interface Classification}: We analyzed the geometric relationship between the ligand and the protein domains to categorize the binding mode. Let $\mathcal{R}_{contact}$ denote the set of protein residues located within a Euclidean distance of 6.0\r{A} from any heavy atom of the ligand. Let $\mathcal{D}(\mathcal{R}_{contact})$ be the set of unique domains to which these contact residues belong. Let $N_{dom}$ represent the total number of structural domains identified within the protein. We categorized complexes into four distinct classes, as illustrated in Figure~\ref{fig:domain_example_flat}:
\begin{enumerate}
    \item Single-Domain: Complexes where $N_{dom}=1$ and $|\mathcal{D}(\mathcal{R}_{contact})|= 1$. These represent the standard baseline for affinity prediction, where the protein consists of a single structural unit containing the binding site.
    \item Single-Domain Binders: Complexes where $N_{dom}\geq2$ and $|\mathcal{D}(\mathcal{R}_{contact})|= 1$. In these cases, the ligand binds exclusively within the pocket of a single domain, despite the presence of adjacent structural units. This category tests the model's ability to discern specific binding sites amidst complex global architectures.
    \item Linker Binders: Complexes where $|\mathcal{D}(\mathcal{R}_{contact})|=\varnothing$. This category encompasses cases where the ligand interacts exclusively with residues residing in non-domain regions (linkers or termini), regardless of whether the protein is single-domain ($N_{dom}=1$) or multi-domain ($N_{dom}\geq2$). This highlights interactions with intrinsically disordered or flexible segments outside defined structural boundaries.
    \item Interface Binders: Complexes where $N_{dom}\geq2$ and $|\mathcal{D}(\mathcal{R}_{contact})|\geq 2$. These ligands bridge the interface between two or more distinct structural domains, representing potential allosteric modulators or stabilizers of specific multi-domain conformations.
\end{enumerate}

Data quality control was enforced through a multistage filtering pipeline. We excluded complexes where the ligand molecular graph could not be chemically validated due to malformed connectivity or coordinated errors. Similarly, protein structures were discarded if they failed parsing or comprised fewer than 10 contacting residues ($\# \mathcal{R}_{contact}<10$), ensuring that only macromolecular targets with sufficient structural context were retained. Finally, to enable domain-aware geometric analysis, we filtered out entries lacking annotated Pfam domains, restricting the dataset to proteins with defined structural units. 

The final curated dataset comprises 16310 protein-ligand complexes. The distribution of samples across the temporal splits (Train, Validation, Test) and the four geometric categories is detailed in Table~\ref{tab:data_stats}.

\begin{table}[h]
\centering
\caption{Distribution of Complexes across Data Splits and Geometric Categories}
\label{tab:data_stats}
\begin{tabular}{l|ccc|c}
\toprule
\textbf{Category} & \textbf{Train} & \textbf{Val} & \textbf{Test} & \textbf{Total} \\
\midrule
Single-Domain & 11342  & 675 & 272 & 12289 \\
Single-Domain Binder & 2246 & 125 & 35 & 2406\\
Linker Binder & 422 & 23 & 8 & 453 \\
Interface Binder & 1059 & 74 & 29 & 1162\\
\midrule
\textbf{Total} & \textbf{15069} & \textbf{897} & \textbf{344} & \textbf{16310} \\
\bottomrule
\end{tabular}
\end{table}

\begin{figure*}[]
\centering
\includegraphics[width=1.0\linewidth]{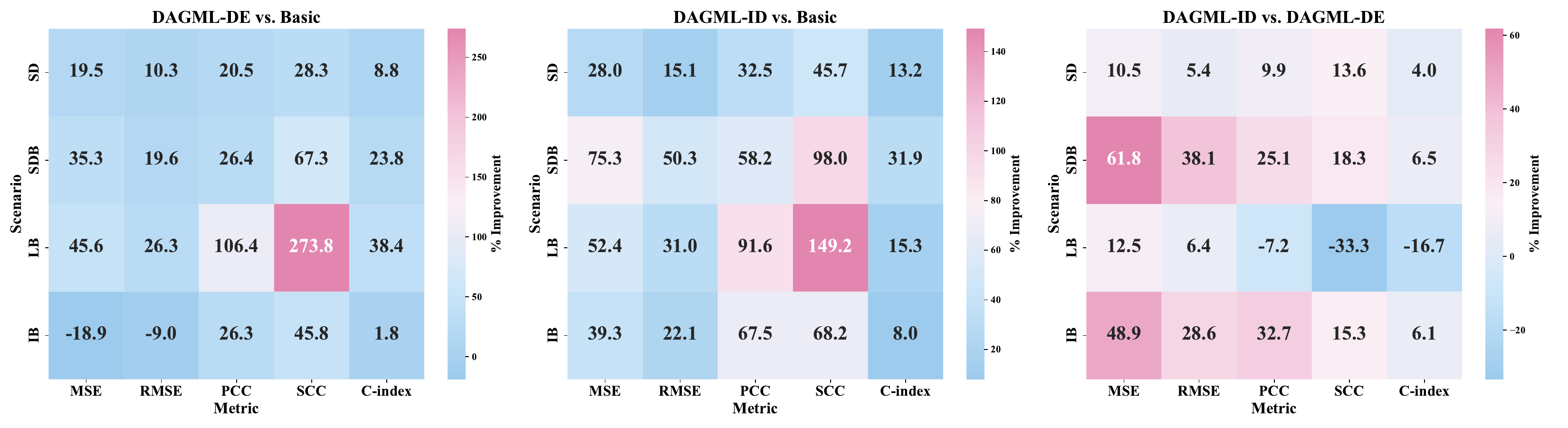}
\caption{Relative performance gain analysis across binding topologies. SD: Single-Domain, SDB: Single-Domain Binder, LB: Linker-Binder, IB: Interface-Binder.}
\label{fig:ablation_types}
\end{figure*}

\section{Experiments and Results}

\subsection{Experimental Setup}
\textbf{Evaluation Metrics:} We assessed model performance using five standard regression metrics following~\cite{Shah2025deepdtagen, Kumar2025casterdta, Huang2025gflearn}: Mean Squared Error (MSE), Root Mean Squared Error (RMSE), Pearson Correlation Coefficient (PCC), Spearman Correlation Coefficient (SCC), and Concordance Index (CI). 

\textbf{Implementation Details:} We implemented DAGML using PyTorch. All models were trained using the Adam optimizer with a learning rate of $1\times10^{-3}$ and a batch size of 128. Training was conducted for a maximum of 50 epochs. To prevent overfitting, we employed an early stopping mechanism that terminates training if the validation loss does not decrease for 5 consecutive epochs. All experiments were conducted on a single 40GB NVIDIA A100 GPU.

\subsection{Impact of Domain-Aware Geometric Modeling}
To evaluate each component's contribution, we conducted a stepwise comparative analysis. We defined three model variants representing increasing levels of structural and semantic granularity.

\begin{enumerate}
    \item Basic: The baseline architecture utilizing the pre-trained ESM-GearNet for protein residue features and MotiL for ligand motif embedding. These features are fused via cross-modal attention and passed to an MLP regressor, without explicit domain definition.
    \item DAGML-DE (Domain Embedding): Extends the Basic model by injecting a learnable domain embedding ($\mathbf{e}_{dom}$ into the protein residue features. It tests whether the model benefits from knowing which functional unit a residue belongs to, without explicitly modeling the spatial interaction between them.
    \item DAGML-ID (Inter-Domain): The complete framework. Building upon DAGML-DE, this variant incorporates the Inter-Domain Interface Graph module. As detailed in the Section~\ref{sec:methodology}, this module performs selective message passing solely across edges connecting residues of distinct domains ($\delta(i)\neq \delta(j)$ within a spatial threshold, explicitly filtering out intra-domain and linker noise.
\end{enumerate}
We evaluated these variants on the multi-domain affinity benchmark, reporting MSE, RMSE, PCC, SCC, C-index. The results are summarized in Table~\ref{tab:main_results}.

\begin{table}[htbp]
\caption{Performance comparison of DAGML variants on the test set. Bold text: the best performance.}
\begin{center}
\begin{tabular}{c|ccccc}
\toprule
Model     & MSE $\downarrow$   & RMSE $\downarrow$  & PCC $\uparrow$   & SCC $\uparrow$   & C-index $\uparrow$ \\ \midrule
Basic     & 2.740 & 1.655 & 0.517 & 0.481 & 0.648   \\ \hline
DAGML-DE & 2.173 & 1.474 & 0.635 & 0.603 & 0.700   \\ \hline
DAGML-ID & \textbf{1.712} & \textbf{1.308} & \textbf{0.726} & \textbf{0.707} & \textbf{0.738}   \\
\bottomrule
\end{tabular}
\label{tab:main_results}
\end{center}
\end{table}

The experimental results provide several insights into the modeling of multi-domain proteins:
\begin{itemize}
    \item \textbf{Impact of Domain Semantics}: The transition from Basic to DAGML-DE yields a substantial performance gain, reducing MSE by approximately 21\% (2.740 to 2.173) and increasing PCC from 0.517 to 0.635. This confirms that standard PLMs and GNNs, which treat the protein as a monolithic sequence or graph, fail to capture the hierarchical nuances of multi-domain topology. By explicitly injecting domain identity, the model can differentiate between structurally similar but functionally distinct units (e.g., catalytic vs. regulatory domains), thereby refining the feature space for affinity prediction.
    \item \textbf{Efficacy of Inter-Domain Message Passing}: The complete DAGML-ID model achieves the highest performance across all metrics, with a further 21\% reduction in MSE compared to the DE variant and a PCC of 0.726. This improvement validates the core hypothesis of our Inter-Domain Interface Graph. Standard GNNs suffer from over-smoothing, where the important geometric signal from domain interfaces is drowned out by the vast number of intra-domain edges. By restricting message passing to the sparse set of edges connecting distinct domains, the ID module acts as a geometric attention mechanism, targeting the clefts and interfaces where allosteric regulation and ligand binding frequently occur.
    \item \textbf{Robustness of Ranking}: The consistent improvement in C-index (0.648$\rightarrow$0.700$\rightarrow$0.738) demonstrates that modeling inter-domain geometry not only improves absolute affinity prediction (MSE) but also enhances the model's ability to correctly rank ligands based on their binding tightness. This is important for virtual screening applications where relative ranking often takes precedence over absolute free energy values.
\end{itemize}

\subsection{Performance Analysis by Binding Scenario}
To understand the source of the performance improvements, we stratified the test set into four distinct binding scenarios as detailed in Section~\ref{sec:domain_annotation}: Single-Domain (SD), Single-Domain Binders (SDB), Linker Binders (LB), and Interface Binders (IB). We visualized the relative performance gains of our proposed models compared to the baseline across these categories in Figure~\ref{fig:ablation_types}.

\subsubsection{Impact of Domain Embeddings (DAGML-DE vs. Basic)}
As shown in Figure~\ref{fig:ablation_types} (Left), introducing explicit domain embeddings yields broad improvements across all categories. Notably, the Linker Binder and Single-Domain Binder categories see the largest relative gains (e.g., 45.6\% and 35.3\% reduction in MSE, respectively). This suggests that explicitly defining structural units helps the model contextualize residue features, particularly in complex multi-domain architectures. 

Interestingly, for Interface Binders, the DE variant improves correlation metrics while error metrics degrade. This indicates that domain semantics alone help the model capture the relative ranking of ligands in complex interfaces, but without explicit modeling of the cleft geometry, the model suffers from calibration errors in absolute affinity prediction.

\subsubsection{Impact of Inter-Domain Message Passing (DAGML-ID vs. Basic)}
The full DAGML-ID model demonstrates dramatic improvements over the baseline, particularly for the most challenging categories (Figure~\ref{fig:ablation_types}, Center). The Interface Binders, which represent the core biological motivation of this work, show a substantial 39.3\% reduction in MSE and a 68.2\% improvement in Spearman correlation. This confirms that the standard GNN's over-smoothing effect is detrimental for interface regions, and that our targeted message passing effectively restores the signal at these critical junctions. Furthermore, the Single-Domain Binder category sees an exceptional 75.3\% reduction in MSE, suggesting that filtering out noise from implicit linkers or unstructured tails sharpens the model's focus on the valid binding pocket even in simpler proteins.

\subsubsection{Dissecting the Gain from Interface Modeling (DAGML-ID vs. DE)}
To isolate the specific contribution of the Inter-Domain GNN, we compared the full model directly against the domain-embedding variant (Figure~\ref{fig:ablation_types}, Right). The results reveal a clear mechanistic advantage:
\begin{enumerate}
    \item Interface Binders: This category shows strong improvement (48.9\% in MSE), which provides strong empirical evidence for our hypotheses: explicitly modeling the spatial proximity of distinct domains allows the network to recognize the unique geometry of inter-domain clefts, a feature that domain embeddings alone cannot capture.
    \item Single-Domain Binders: This category benefits most significantly with a 61.8\% reduction in MSE and a 18.3\% improvement in SCC. The model effectively discerns the active pocket from adjacent decoy domains without requiring extensive cross-domain message passing.
    \item Linker Binders: Conversely, for this category, the ID model reduces error metrics (12.5\% in MSE) but degrades ranking metrics (16.7\% in C-index). This partially confirms our design hypothesis: the ID module explicitly filters out linker residues to focus on rigid domain interfaces. For ligands that bind exclusively to these flexible linkers, this filtering removes critical geometric signal, causing the model to revert to conservative, mean-centric predictions that lower absolute error but fail to capture specific binding trends.
\end{enumerate}

\begin{figure}[htbp]
\centering
\includegraphics[width=0.9\columnwidth]{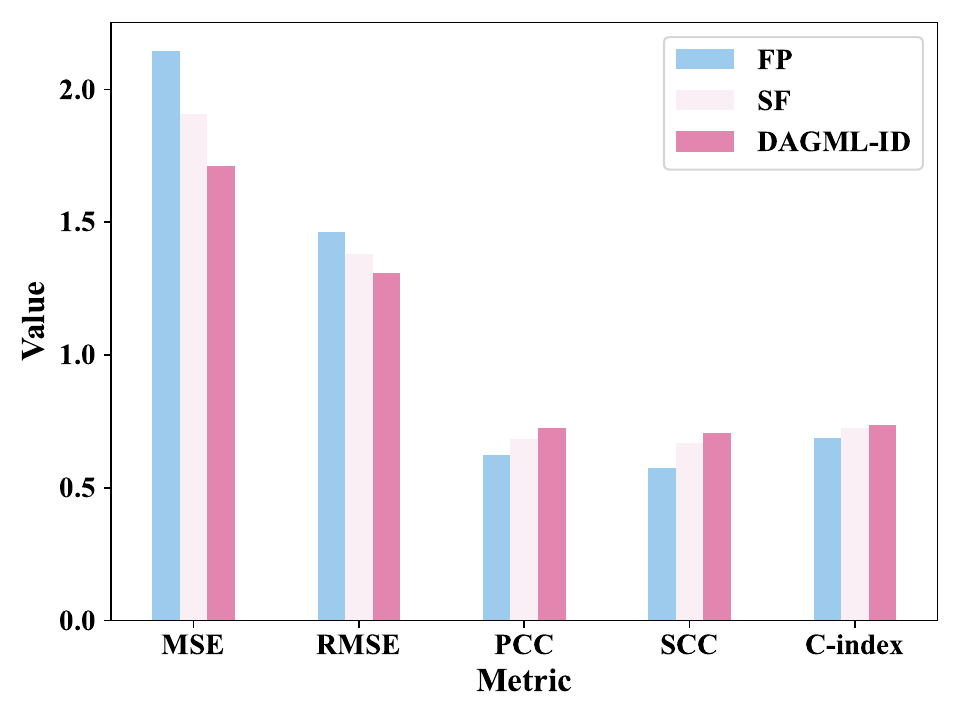}
\caption{Impact of feature representation modalities. FP (blue): replacing Molecular Motif Learning with Morgan Fingerprints; SF (pink): sequential feature replacing ESM-GearNet with ESM-2; DAGML-ID (rose): full structure-and motif-aware framework.}
\label{fig:ablation_feat}
\end{figure}

\subsection{Impact of Feature Representation Modalities}
We conducted an ablation study involving two input feature variants of our best-performing models (DAGML-ID). The comparative results across five metrics are shown in Figure~\ref{fig:ablation_feat}.

\subsubsection{Ligand Representation (FP vs. DAGML-ID)}
To assess the contribution of the Molecular Motif Learning (MotiL) encoder, we replaced the motif-aware embeddings with standard Morgan Fingerprints (FP), a widely used fixed-length bit vector representation for small molecules. The results indicate a performance drop with MSE increasing from 1.712 to 2.145 and the PCC dropping from 0.726 to 0.623. This confirms that standard fingerprints fail to capture the rich semantic information that is important for the cross-modal attention mechanism to align ligand with specific protein domains.

\subsubsection{Protein Representation (SF vs. DAGML-ID)}
To evaluate the importance of geometric structural information, we replaced the structure-aware ESM-GearNet encoder with the sequence-only ESM-2 model feature (SF). While the sequence-only model outperforms the fingerprint-based variant, it still lags behind the full DAGML-ID framework. The RMSE increased from 1.308 to 1.381, and the SCC dropped from 0.707 to 0.669. The performance indicates that while evolutionary information is powerful, it is insufficient for resolving the complex spatial arrangements of multi-domain interface. Explicit geometric encoding is required to accurately model the specific inter-domain feature.

\begin{figure}[htbp]
\centering
\includegraphics[width=0.85\columnwidth]{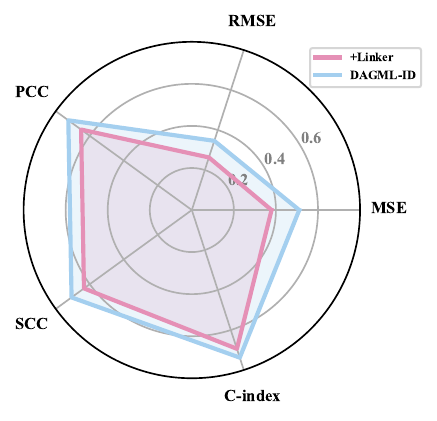}
\caption{Impact of linker filtering on geometric noise. +Linker (purple) represents that the restrictions on flexible linker residues were relaxed during graph construction. MSE and RMSE are inverted so that larger areas indicate better performance.}
\label{fig:ablation_linker}
\end{figure}

\subsubsection{Impact of Linker Filtering on Geometric Noise}
A key design choice in our Inter-Domain GNN is the explicit exclusion of residues belonging to flexible linkers (i.e., unstructured regions connecting annotated domains) during graph construction. To validate this design, we conducted an ablation study where this constraint was relaxed, allowing passing to occur between domains and linker regions (denoted as +Linker). The results are visualized in Figure~\ref{fig:ablation_linker}.

For the radar plot analysis, the error metrics (MSE and RMSE) were inverted and normalized. As illustrated by the plot, the proposed DAGML-ID completely encompasses the +Linker variant. Quantitatively, the MSE increased by 26.9\% (from 1.712 to 2.173), while the PCC dropped from 0.726 to 0.651. This degradation supports our hypothesis that flexible linkers introduce geometric noise. Unlike structure domains, linker coordinates in static crystal structures often exhibit high B-factors or are arbitrarily modeled due to their intrinsic disorder. By permitting message passing through these regions, the graph topology becomes saturated with non-informative edges.

\section{Limitations and Future Work}
Filtering linker binders creates a trade-off: it sharpens signals for rigid interfaces but degrades ranking performance for linker-bound ligands by removing geometric context. Future work will address this limitation using dynamic graph representations, such as adaptive attention weighted by pLDDT scores. This will allow the model to selectively retain signal from stable linker interactions while suppressing disorder-induced noise.

\section{Conclusion}
In this work, we presented DAGML, a unified framework for predicting protein-ligand affinity that moves beyond the monolithic treatment of protein structures. By explicitly encoding domain semantics and modeling the sparse geometry of inter-domain interfaces, our model overcomes the over-smoothing limitations of standard GNNs in complex architectures. We also introduce the first geometrically stratified multi-domain affinity benchmark. By classifying complexes based on binding topology, this dataset provides a benchmark for evaluating domain-aware algorithms. The proposed method was validated on a topologically stratified framework, demonstrating good performance not only in overall accuracy but specifically in targeting complex interface binding sites. Our analysis confirms that while sequence-based features provide a necessary foundation, the geometric resolution of domain clefts and the filtering of flexible linker noise are essential for robust predictions in multi-domain systems. DAGML thus offers a computational tool for targeting the expansive and often undrugged space of multi-domain proteins.

\bibliographystyle{IEEEtran}
\bibliography{refs}

\end{document}